\documentclass[aps,twocolumn,showpacs,showkeys,preprintnumbers,amsmath]{revtex4}

\allowdisplaybreaks[1]
\usepackage{amssymb}
\usepackage{graphicx}
\newcommand{\nc}{\newcommand}
\nc{\rnc}{\renewcommand}
\nc{\dmo}{\DeclareMathOperator}

\nc{\nojour}[3]{\textbf{#1}, #2 (#3)}
\nc{\arXiv}[3]{arXiv:#1.#2v#3}
\nc{\ibid}[3]{{\em ibid.} \textbf{#1}, #2 (#3)}
\nc{\EPJB}[3]{Eur.\ Phys.\ J.\ B \textbf{#1}, #2 (#3)}
\nc{\EPL}[3]{Europhys.\ Lett.\ \textbf{#1}, #2 (#3)}
\nc{\JPA}[3]{J.\ Phys.\ A \textbf{#1}, #2 (#3)}
\nc{\JPAMG}[3]{J.\ Phys.\ A: Math.\ Gen.\ \textbf{#1}, #2 (#3)}
\nc{\JPAMT}[3]{J.\ Phys.\ A: Math.\ Theor.\ \textbf{#1}, #2 (#3)}
\nc{\JSMTE}[2]{J.\ Stat.\ Mech.: Theory Exp.\ (#1) #2}
\nc{\JSP}[3]{J.\ Stat.\ Phys.\ \textbf{#1}, #2 (#3)}
\nc{\PAA}[3]{Physica \textbf{#1}, #2 (#3)}
\nc{\PL}[3]{Phys.\ Lett.\ \textbf{#1}, #2 (#3)}
\nc{\PR}[3]{Phys.\ Rev.\ \textbf{#1}, #2 (#3)}
\nc{\PRB}[3]{Phys.\ Rev.\ B \textbf{#1}, #2 (#3)}
\nc{\PRE}[3]{Phys.\ Rev.\ E \textbf{#1}, #2 (#3)}
\nc{\PRL}[3]{Phys.\ Rev.\ Lett.\ \textbf{#1}, #2 (#3)}
\nc{\RMP}[3]{Rev.\ Mod.\ Phys.\ \textbf{#1}, #2 (#3)}
\nc{\ZPB}[3]{Z.\ Phys.\ B: Condens.\ Matter \textbf{#1}, #2 (#3)}

\dmo{\sgn}{sgn}
\dmo{\diag}{diag}
\dmo{\arsinh}{arsinh}

\nc{\bse}{\begin{subequations}}
\nc{\ese}{\end{subequations}}

\nc{\ts}{\textstyle}
\nc{\unit}{\pmb{\openone}}
\nc{\f}{\frac}
\nc{\fr}[2]{{\ts\f{#1}{#2}}}

\nc{\al}{\alpha}
\nc{\be}{\beta}
\nc{\vp}{\varphi}
\rnc{\th}{\theta}
\nc{\De}{\Delta}

\nc{\Abar}{\bar{\mathbf{A}}}
\nc{\Rb}{\mathbf{R}}
\nc{\Xib}{\mathbf{\Xi}}

\nc{\Fcal}{{\cal F}}

\nc{\Tc}{T_\text{c}}

\nc{\xig}{\xi_>}
\nc{\xik}{\xi_<}
\nc{\xipz}{\xi_{+,0}}
\nc{\ximz}{\xi_{-,0}}
\nc{\xipmz}{\xi_{\pm,0}}

\nc{\bc}{{bc}}

\nc{\Jd}{J_\text{d}}

\begin{document}

\bibliographystyle{apsrev}

\title{Anisotropy and universality in finite-size scaling:\\
Critical Binder cumulant of a two-dimensional Ising model}
\author{Boris Kastening}
\email[Email address: ]{bkastening@matgeo.tu-darmstadt.de}
\affiliation{Institute for Materials Science,
Technische Universit\"at Darmstadt, D-64287 Darmstadt, Germany}
\date{\today}

\begin{abstract}
We reanalyze transfer-matrix and Monte Carlo results for the critical
Binder cumulant $U^*$ of an anisotropic two-dimensional Ising model on a
square lattice in a square geometry with periodic boundary conditions.
Spins are coupled between nearest-neighboring sites and between
next-nearest-neighboring sites along one of the lattice diagonals.
We find that $U^*$ depends only on the asymptotic critical long-distance
features of the anisotropy, irrespective of its realization through
ferromagnetic or antiferromagnetic next-nearest-neighbor couplings.
We modify an earlier renormalization-group calculation to obtain a
quantitative description of the anisotropy dependence of $U^*$.
Our results support our recent claim towards the validity of universal
finite-size scaling for critical phenomena in the presence of a weak
anisotropy.
\end{abstract}



\maketitle

Universality is a key concept in the theory of critical phenomena for
bulk and confined systems; for reviews, see, for example,
\cite{Fi74,Pr90,PrAhHo91}.
Within a given bulk universality class, critical exponents, certain
critical amplitude ratios, and the critical behavior of thermodynamic
functions are identical and independent of the specific microscopic
realization.
For instance, the bulk correlation length in the asymptotic critical
domain, that is, for asymptotically small positive or negative
$t\equiv(T-\Tc)/\Tc$, is, for isotropic systems, described by
\begin{align}
\label{xi.crit}
\xi&=\xipmz|t|^{-\nu}, & T&\gtrless\Tc,
\end{align}
where $\Tc$ is the bulk critical temperature, $\nu>0$ is a universal
critical exponent, $\xipmz$ are nonuniversal critical amplitudes, and
$R_\xi\equiv\xipz/\ximz$ is a universal critical amplitude ratio.

Besides critical exponents, the thermodynamic functions describing the
asymptotic critical domain are universal if one allows for adjusting
of only two amplitudes.
Consider, for example, for small $t$ and small field $h$ conjugate to the
order parameter, the part $f_\text{b,s}$ of the bulk free energy density
that becomes singular at the critical point $t,h=0$.
$f_\text{b,s}$ is asymptotically described by universal scaling functions
$W^\pm$ for $t\gtrless0$ according to \cite{PrFi84}
\begin{align}
\label{f.bs}
\be f_\text{b,s}(t,h)
&=
A_1|t|^{2-\al} W^\pm(A_2h|t|^{-\De}),
\end{align}
with $\be\equiv1/(k_\text{B}T)$, with universal critical exponents $\al$
and $\De$, and where $A_1$ and $A_2$ are nonuniversal
amplitudes, that is, they differ between different systems within the
universality class under consideration.

Universality also extends to situations, where the system under
consideration is confined on a length scale $L$ that is large compared to
all microscopic length scales, such as lattice constants.
If the singular part of the free energy density of the system under
consideration exhibits scaling, its asymptotic critical form may be
written as \cite{PrFi84}
\begin{align}
\label{f.s}
\be f_\text{s}(t,h,L)
&=
L^{-d}\Fcal(C_1tL^{1/\nu},C_2hL^{\De/\nu}),
\end{align}
where the nonuniversal constants $C_1$ and $C_2$ are universally related
to the constants $A_1$ and $A_2$ in (\ref{f.bs}).
For given shape and boundary conditions, the function $\Fcal$ is
universal.

Equations (\ref{xi.crit})--(\ref{f.s}) were stated for isotropic critical
systems, that is, where the bulk asymptotic near-critical correlation
length is isotropic.
Here we are interested in the fate of universal finite-size scaling in
the presence of a weak anisotropy, that is, where the amplitudes $\xipmz$
depend on the direction, but $\nu$ does not (for strong anisotropies,
where even $\nu$ depends on the direction, see, for example, \cite{To07}
and references therein).
Representatives of spatially anisotropic systems in the context of
critical phenomena are, for example, magnetic materials, alloys,
superconductors \cite{Sc08}, and solids with structural phase transitions
\cite{BrCo81,Sa92}.

Most of the literature on critical phenomena has focused on the isotropic
case.
In Refs.~\cite{InNiWa86,NiBl83,KiPe87,KaLaTu97,Ca83,BaPePe84,Yu97,Hu02,
SeSh05,SeSh09}, investigations were carried out on a number of specific
anisotropic systems, but no general picture of universality for weakly
anisotropic systems in restricted geometries was suggested.
The more recent publications \cite{ChDo04,Do06,Do08,DiCh09,Ka12} have
attempted to clarify the situation from a more general point of view.
References~\cite{ChDo04,Do06,Do08} concluded that the dependence of some
bulk amplitude ratios, scaling functions, and the critical Binder cumulant
on the parameters describing the anisotropy indicates a violation of
universality.
In Ref.~\cite{DiCh09}, it was suggested that universality should be defined
only after relating the system under consideration to an isotropic system
by means of a shear transformation (see \cite{InNiWa86} for such a
suggestion in the context of the two-dimensional Ising model).

In contrast, the present author suggested in \cite{Ka12} that quantities
that are universal in the isotropic case remain universal in the presence
of a weak anisotropy, if their list of arguments is augmented by the
parameters describing the asymptotic critical long-distance features of
the anisotropy.
Universality then implies that the quantity under consideration should
exhibit no dependence on the particular microscopic realization of these
features.
Rather, it should depend universally on the anisotropy, that is, in a way
that is identical for all members of the respective bulk universality
class.
Explicit support for this claim was provided in \cite{Ka12} through
results for free energy scaling functions of the two-dimensional Ising
model on infinite strips.
These examples, however, exhibited the simplification that a shear
transformation exactly related the anisotropic systems to their isotropic
counterparts with identical boundary conditions and geometry.
Accordingly, the scaling functions for the anisotropic case could be
expressed in terms of the isotropic scaling functions, and the
consequences of the anisotropy could be interpreted to be of a mere
geometric nature.
In this work, we consider a situation where this simplification is absent.

Consider weakly anisotropic critical bulk systems in $d$ dimensions.
Their asymptotic long-distance correlations are described by a correlation
length ellipsoid that may be represented by a symmetric positive definite
matrix $\Xib$.
This is diagonalized by a rotation matrix $\Rb$, so that
\begin{align}
\Rb\Xib\Rb^{-1}=\diag(\xi_1^2,\xi_2^2,\ldots,\xi_d^2).
\end{align}
The $\xi_i$ are the asymptotic near-critical bulk correlation lengths
along the principal axes of the ellipsoid.
The scale-free matrix \cite{Ka12}
\begin{align}
\label{Ab}
\Abar\equiv(\det\Xib)^{-1/d}\Xib
\end{align}
is normalized to have $\det\Abar=1$ and describes the shape and
orientation of the ellipsoid.
Since $\nu$ is unique by assumption, $\Abar$ is independent of $t$ within
the asymptotic critical domain.
It may be parametrized by $d{-}1$ correlation length ratios and
$d(d{-}1)/2$ rotation angles.
According to \cite{Ka12}, universal quantities may receive an additional
dependence on $\Abar$, but are independent of its particular microscopic
realization.
This implies that the functions $W^\pm(x)$ and $\Fcal(y,z)$ on the
right-hand sides of Eqs.~(\ref{f.bs}) and (\ref{f.s}), respectively, have
to be replaced by universal functions $W^\pm(x,\Abar)$ and
$\Fcal(y,z,\Abar)$, which describe the isotropic situation as a special
case $\Abar=\unit$.

The use of a normalized matrix $\Abar$ to describe the anisotropy of a
model was suggested in \cite{ChDo04,Do06,Do08}.
However, while $\Abar$ from (\ref{Ab}) is defined through the physical
correlation lengths in the asymptotic critical domain, explicit versions
of $\Abar$ in \cite{ChDo04,Do06,Do08} were obtained by expanding the
Hamiltonian under consideration in small wave numbers $k$ through order
$k^2$.
While, for standard $\vp^4$ field theory, these definitions coincide due
to an exact mapping between anisotropic and isotropic bulk Hamiltonians
\cite{ChDo04}, this procedure in general does not, for lattice models,
lead to the same $\Abar$ as defined through the physical correlation
lengths.
This will be demonstrated explicitly for the two-dimensional Ising model
below.
The matrix $\Abar$ used in \cite{ChDo04,Do06,Do08} is thus not universally
related to the matrix defined in (\ref{Ab}). This explains why the author
of \cite{Do06,Do08} arrived at the conclusion that the anisotropy
effects exhibit ``a kind of restricted universality'' that only holds for
a subclass of systems within a given universality class.
Here we give support to our claim from \cite{Ka12} that the use of the
matrix $\Abar$ as defined in (\ref{Ab}) leads to unrestricted universality
in the traditional sense.

\begin{figure}[t]
\begin{center}
$\underset{\mbox{(a)}\rule{0pt}{15pt}}{\quad
\includegraphics[width=4cm,angle=0]{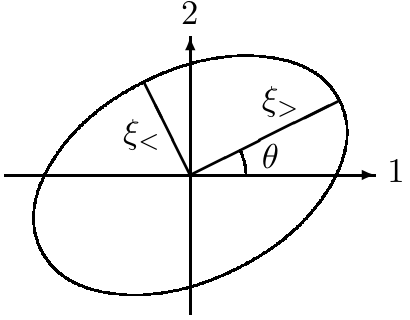}}$
\qquad\qquad
$\underset{\mbox{(b)}\rule{0pt}{15pt}}{
\includegraphics[width=2cm,angle=0]{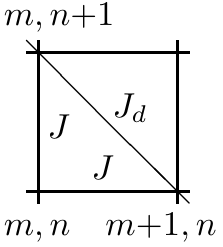}}$
\end{center}
\caption{\label{ellipse.lattice}
(a) Bulk near-critical correlation length ellipse for two-dimensional
models with an angle of inclination $\th$ and largest and smallest
correlation lengths $\xig$ and $\xik$, respectively.
For the two-dimensional Ising model, the ``$1$'' and ``$2$'' directions
are chosen as the $m$ and $n$ principal directions, respectively, of the
square lattice.
(b) Ising model square lattice with nearest-neighbor couplings $J$ and
next-nearest-neighbor couplings $\Jd$ on one of the diagonals.
}
\end{figure}

For $d=2$-dimensional models, the ellipsoid reduces to an ellipse
\cite{Va76}, whose shape may be characterized by the ratio
$r\equiv\xi_</\xi_>$ of the smallest and largest correlation lengths (thus
$0<r\leq1$) and an angle $\th$ describing its orientation (we select the
convention $-\pi/2<\th\leq\pi/2$).
If $\th$ is chosen to be the inclination of the direction of the largest
correlation length with respect to the ``$1$'' direction, see
Fig.~\ref{ellipse.lattice}(a), the explicit form of $\Abar$ for $d=2$ is
\cite{Ka12}
\begin{align}
\label{Abar}
\Abar_2=
\begin{pmatrix}
r\sin^2\th+r^{-1}\cos^2\th & \fr{1}{2}(r^{-1}{-}r)\sin2\th \\
\fr{1}{2}(r^{-1}{-}r)\sin2\th & r\cos^2\th+r^{-1}\sin^2\th
\end{pmatrix}.
\end{align}
Therefore, universal quantities may receive an additional dependence on
the variables $r$ and $\th$ as compared to the isotropic case.

Consider a two-dimensional Ising model in a square $L\times L$ geometry
on a square  lattice with lattice constant $a$ and ferromagnetic couplings
$J>0$ of neighboring spins and couplings $\Jd$ of next-nearest-neighboring
spins in the direction of only one of the diagonals, see
Fig.~\ref{ellipse.lattice}(b).
With $L=Na$, the Hamiltonian of this model reads
\begin{align}
\label{H}
H
=
-\sum_{m,n=1}^Ns_{m,n}\big[&J(s_{m+1,n}+s_{m,n+1})
+\Jd s_{m+1,n+1}\big],
\end{align}
with identifications of coordinates $N+1$ and $1$ in both principal
lattice directions (i.e., periodic boundary conditions) and where
$s_{m,n}=\pm1$ is the Ising spin at the site $(m,n)$.
We only consider couplings $\Jd>-J$ that allow for a ferromagnetic bulk
phase transition at a temperature $\Tc>0$, given by \cite{Ho50,BeHu93}
\begin{align}
\label{crit2}
\sinh^2(2\be_\text{c}J)+2\sinh(2\be_\text{c}J)\sinh(2\be_\text{c}\Jd)=1.
\end{align}
We choose $\th$ to be the inclination of the direction of the largest
correlation length with respect to one of the lattice axes.
Using results for the general triangular lattice \cite{InNiWa86,Ka12}, we
obtain, at the critical point,
\begin{align}
\label{r.th.crit}
r&=[\sinh(2\be_\text{c}J)]^{\pm1}, & \th&=\mp\pi/4
&\text{for~~} \Jd&\gtrless0 .
\end{align}
Due to the symmetry of the $L\times L$ geometry, no dependence
of universal critical quantities on the sign of $\th$ in (\ref{r.th.crit})
is possible.
Thus, compared to the isotropic case, universal quantities may receive an
additional dependence only on $r$.
Note that for general $r<1$, a shear transformation to an isotropic
system causes the periodicity of the boundary conditions to be no longer
in mutually perpendicular directions and therefore no transformation to an
isotropic system with identical geometry and boundary conditions to the
original system is possible.
Thus the consequences of the anisotropy cannot be interpreted to be of a
mere geometric nature.

Here we test anisotropic universality for the critical Binder cumulant.
The Binder cumulant \cite{Bi81} is a measure of the order parameter
distribution.
It may be used to locate the phase transition for a given model from the
intersection of the cumulant for different system sizes or to determine
the critical exponent $\nu$ and therefore the universality class.
For our system, the Binder cumulant is defined by
\begin{align}
U(T,L)&\equiv1-\langle M^4\rangle/(3\langle M^2\rangle^2),
\end{align}
with the magnetization per spin $M\equiv N^{-2}\sum_{m,n=1}^Ns_{m,n}$
and its second and fourth moments $\langle M^2\rangle$ and
$\langle M^4\rangle$, respectively.
We are interested in the critical thermodynamic limit
$U^*=\lim_{L\to\infty}U(\Tc,L)$.
For given universality class, geometric system shape (taken to the
thermodynamic limit), and boundary conditions, $U^*$ is believed to be
universal, that is, to have a unique value in the isotropic case,
independent of the underlying lattice \cite{JaKaVi94,Se07} and the spin
value \cite{NiBr88}.
According to our considerations above, we expect $U^*$ to become a
universal function $U^*(r)$ for the anisotropic case discussed here, that
is, if we switch on $\Jd$.

\begin{figure}[tH]
\begin{center}
\includegraphics[width=8.5cm,angle=0]{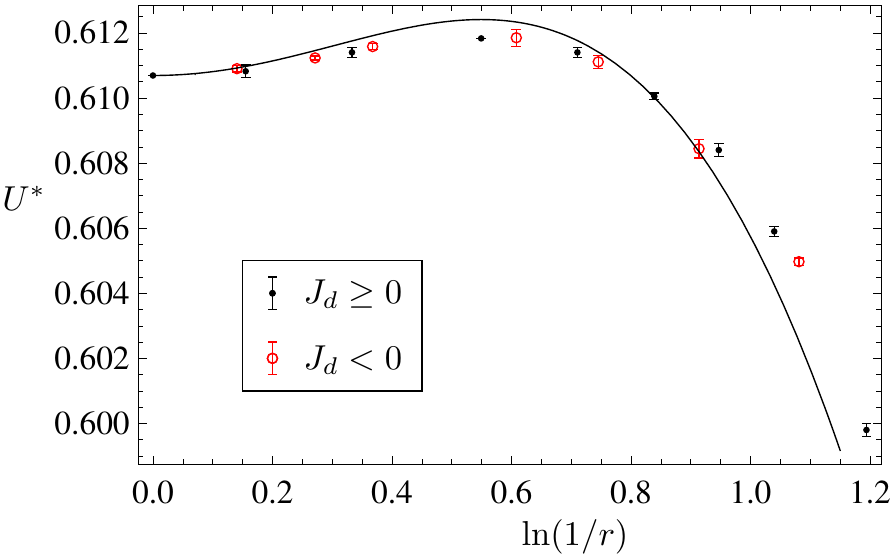}
\end{center}
\caption{\label{caniso}
(Color online)
Transfer-matrix results from \cite{KaBl93} and Monte Carlo results from
\cite{SeSh05,SeSh09} for the Binder cumulant $U^*$ as a function of
$\ln(1/r)$ \cite{MC_comment}.
Solid line: $d{=}3$ renormalization-group estimate for the relative change
of $U^*$ for $d{=}2$ away from its isotropic value at $r=1$.}
\end{figure}

The most precise related results were obtained in \cite{KaBl93} by means
of a transfer-matrix method and imply
\bse
\label{U.KaBl}
\begin{align}
\label{U.KaBl.sq}
U^*(1)&=0.610\,690\,1(5), & \Jd&=0,\\
\label{U.KaBl.tr}
U^*(1/\sqrt{3})&=0.611\,827\,74(2), & \Jd&=J.
\end{align}
\ese
Using Monte Carlo techniques, precision results have also been obtained
for other values of $\Jd$, notably for both $\Jd>0$ \cite{SeSh05} and
$\Jd<0$ \cite{SeSh09}.
All of these results were displayed in Fig.~4 of \cite{SeSh09} as a
function of $s\equiv1/[1+(J/\Jd)]$.
Using the relation of $r$ and $s$ provided by the definition of $s$
together with Eqs.~(\ref{crit2}) and (\ref{r.th.crit}), we have plotted
all available data points from Refs.~\cite{KaBl93,SeSh05,SeSh09} as a
function of $\ln(1/r)$ in Fig.~\ref{caniso}.
Without universality, one would expect the Binder cumulant for
ferromagnetic and antiferromagnetic couplings $\Jd$ to be described by two
separate functions $U_\text{fm}^*(r)$ and $U_\text{afm}^*(r)$ that
coincide for the isotropic case $r=1$ (where $\Jd=0$), but may differ for
other values of $r$.
Note, however, that all data points appear to be described by a single
function $U^*(r)$, irrespective of the sign of $\Jd$, that is, for
microscopic realizations of $r$ through both ferromagnetic and
antiferromagnetic couplings $\Jd$.
This is the central observation of this work, giving strong support to
the conjectured universality of weakly anisotropic critical systems.

We conclude by discussing the use of the renormalization group (RG) to
describe the anisotropy dependence of $U^*$.
In \cite{Do08}, the RG in three dimensions was used to describe the
anisotropy dependence of $U^*$ in two dimensions.
A two-dimensional anisotropy matrix
$\Abar_2(s)=(1-s^2)^{-1/2}
\left(\begin{smallmatrix}
1 & s \\
s & 1
\end{smallmatrix}\right)$
with $s$ from above was incorporated into a block-diagonal
$3\times3$ matrix $\Abar_3(s)$ to estimate the $s$ dependence of $U^*$.
The result provided a good approximation to the available data from
\cite{KaBl93,SeSh05} for $\Jd\geq0$.
However, later Monte Carlo data for $\Jd<0$ were not in agreement with the
RG results, unless $\Jd$ is small \cite{SeSh09}.

The reason for this discrepancy is easy to identify.
The matrix $\Abar_2(s)$ does not provide the correct description of the
asymptotic shape of the correlation length ellipse in contrast to
\begin{align}
\label{Ab2.r}
\Abar_2(r)
&=
\f{1}{2}
\begin{pmatrix}
r+r^{-1} & \mp(r^{-1}-r)\\
\mp(r^{-1}-r) & r+r^{-1} 
\end{pmatrix},
\end{align}
obtained by setting $\th=\mp\pi/4$ in (\ref{Abar}) for $\Jd\gtrless0$.
While the deviation is small for positive $\Jd$, it becomes large for
negative $\Jd$, unless $\Jd$ is small.
Using $\Abar_2(r)$ instead, a natural choice for the appropriate
three-dimensional matrix is
\begin{align}
\label{Ab3.r}
\Abar&=\Abar_3(r)=
\begin{pmatrix}
\Abar_2(r) & \begin{matrix}0\\0\end{matrix}  \\
\begin{matrix}0&0\end{matrix} & 1
\end{pmatrix},
\end{align}
which localizes the anisotropy within the first two dimensions.
Plugging our choice of $\Abar_3(r)$ into the RG result for
$U_\text{RG}^*(r)\equiv U(0,\Abar)$, provided in Eq.~(7.8) of \cite{Do08},
and following Ref.~\cite{SeSh09} in normalizing the resulting function
$U_\text{RG}^*(r)$ by plotting $U_\text{RG}^*(r)U^*(1)/U_\text{RG}^*(1)$
with $U^*(1)$ from (\ref{U.KaBl.sq}), we obtain the solid line in
Fig.~\ref{caniso}, valid for both signs in (\ref{Ab2.r}) and thus for both
ferromagnetic and antiferromagnetic $\Jd$.
We conclude that the three-dimensional RG calculation of \cite{Do08}
results in a good description of the relative anisotropy dependence of the
two-dimensional critical Binder cumulant, provided the anisotropy is
described by the appropriate matrix $\Abar_2(r)$ from (\ref{Ab2.r}) and
$\Abar_2(r)$ is naturally embedded into $\Abar_3(r)$ as in (\ref{Ab3.r}).

In summary, we considered the critical Binder cumulant $U^*$ of a
two-dimensional Ising model on a square lattice in a square geometry with
periodic boundary conditions and nearest-neighbor couplings.
Additional couplings on one of the lattice diagonals allowed us to adjust
a variable diagonal anisotropy of the correlation lengths in the
asymptotic critical domain.
We found that $U^*$ depends universally on its argument $r$,
defined as the ratio of the smallest and largest bulk correlation length,
independently of the particular realization of $r$ through ferromagnetic
or antiferromagnetic diagonal couplings.
Observing the true physical asymptotic long-distance behavior of the
anisotropy, represented by the asymptotic critical correlation length
ellipse, we repeated Dohm's renormalization group calculation and found
good quantitative agreement of the relative anisotropy dependence of $U^*$
with precision transfer-matrix and Monte Carlo data.
Our results support the validity of universal finite-size scaling for
critical phenomena in the presence of a weak anisotropy.
The ideas presented here and in \cite{Ka12} may be used to analyze other
quantities in weakly anisotropic systems, for example, those (besides the
Binder cumulant) for which nonuniversality was claimed in
\cite{ChDo04,Do06,Do08}.

\section*{Acknowledgments}
The author is grateful to W.~Selke for useful communications.
He thanks W.~Selke and L.N.~Shchur for providing their original Monte
Carlo data of Refs.~\cite{SeSh05,SeSh09}.


\begin{thebibliography}{9}
\bibitem{MC_comment}
We show the Monte Carlo data provided to us by W.~Selke and L.N.~Shchur.
For $s=0.6$ and $s=0.6666$, corresponding to $\ln(1/r)\approx0.710$ and $\ln(1/r)\approx0.839$, respectively, their error bars are somewhat larger
than those plotted in Fig.~4 of \cite{SeSh09}.
\end{thebibliography}
\end{document}